\documentclass[12pt,a4]{article}
\usepackage{amsfonts}

\makeatletter
\newcount\@hour\newcount\@minute\chardef\@x10\chardef\@xv60
\def\tcitime{
\def\@time{%
  \@minute\time\@hour\@minute\divide\@hour\@xv
  \ifnum\@hour<\@x 0\fi\the\@hour:%
  \multiply\@hour\@xv\advance\@minute-\@hour
  \ifnum\@minute<\@x 0\fi\the\@minute
  }}%
\def\QCTOpt[#1]#2{%
  \def\QCTOptB{#1}
  \def\QCTOptA{#2}
}
\def\QCTNOpt#1{%
  \def\QCTOptA{#1}
  \let\QCTOptB\empty
}
\def\Qct{%
  \@ifnextchar[{%
    \QCTOpt}{\QCTNOpt}
}
\def\QCBOpt[#1]#2{%
  \def\QCBOptB{#1}
  \def\QCBOptA{#2}
}
\def\QCBNOpt#1{%
  \def\QCBOptA{#1}
  \let\QCBOptB\empty
}
\def\Qcb{%
  \@ifnextchar[{%
    \QCBOpt}{\QCBNOpt}
}
\def\PrepCapArgs{%
  \ifx\QCBOptA\empty
    \ifx\QCTOptA\empty
      {}%
    \else
      \ifx\QCTOptB\empty
        {\QCTOptA}%
      \else
        [\QCTOptB]{\QCTOptA}%
      \fi
    \fi
  \else
    \ifx\QCBOptA\empty
      {}%
    \else
      \ifx\QCBOptB\empty
        {\QCBOptA}%
      \else
        [\QCBOptB]{\QCBOptA}%
      \fi
    \fi
  \fi
}
\newcount\GRAPHICSTYPE
\GRAPHICSTYPE=\z@
\def\GRAPHICSPS#1{%
 \ifcase\GRAPHICSTYPE%\GRAPHICSTYPE=0
   \special{ps: #1}%
 \or%\GRAPHICSTYPE=1
   \special{language "PS", include "#1"}%
 \fi
}%
\def\graffile#1#2#3#4{%
    \leavevmode
    \raise -#4 \BOXTHEFRAME{%
        \hbox to #2{\raise #3\hbox{\null #1}}}%
}%
\def\draftbox#1#2#3#4{%
 \leavevmode\raise -#4 \hbox{%
  \frame{\rlap{\protect\tiny #1}\hbox to #2%
   {\vrule height#3 width\z@ depth\z@\hfil}%
  }%
 }%
}%
\newcount\draft
\draft=\z@

\newif\ifwasdraft
\wasdraftfalse

\def\GRAPHIC#1#2#3#4#5{%
 \ifnum\draft=\@ne\draftbox{#2}{#3}{#4}{#5}%
  \else\graffile{#1}{#3}{#4}{#5}%
  \fi
 }%
\def\addtoLaTeXparams#1{%
    \edef\LaTeXparams{\LaTeXparams #1}}%
\newif\ifBoxFrame \BoxFramefalse
\newif\ifOverFrame \OverFramefalse
\newif\ifUnderFrame \UnderFramefalse

\def\BOXTHEFRAME#1{%
   \hbox{%
      \ifBoxFrame
         \frame{#1}%
      \else
         {#1}%
      \fi
   }%
}

\def\doFRAMEparams#1{\BoxFramefalse\OverFramefalse\UnderFramefalse\readFRAME
params#1\end}%
\def\readFRAMEparams#1{%
 \ifx#1\end%
  \let\next=\relax
  \else
  \ifx#1i\dispkind=\z@\fi
  \ifx#1d\dispkind=\@ne\fi
  \ifx#1f\dispkind=\tw@\fi
  \ifx#1t\addtoLaTeXparams{t}\fi
  \ifx#1b\addtoLaTeXparams{b}\fi
  \ifx#1p\addtoLaTeXparams{p}\fi
  \ifx#1h\addtoLaTeXparams{h}\fi
  \ifx#1X\BoxFrametrue\fi
  \ifx#1O\OverFrametrue\fi
  \ifx#1U\UnderFrametrue\fi
  \ifx#1w
    \ifnum\draft=1\wasdrafttrue\else\wasdraftfalse\fi
    \draft=\@ne
  \fi
  \let\next=\readFRAMEparams
  \fi
 \next
 }%
\def\IFRAME#1#2#3#4#5#6{%
      \bgroup
      \let\QCTOptA\empty
      \let\QCTOptB\empty
      \let\QCBOptA\empty
      \let\QCBOptB\empty
      #6%
      \parindent=0pt%
      \leftskip=0pt
      \rightskip=0pt
      \setbox0 = \hbox{\QCBOptA}%
      \@tempdima = #1\relax
      \ifOverFrame
          \typeout{This is not implemented yet}%
          \show\HELP
      \else
         \ifdim\wd0>\@tempdima
            \advance\@tempdima by \@tempdima
            \ifdim\wd0 >\@tempdima
               \textwidth=\@tempdima
               \setbox1 =\vbox{%
                  \noindent\hbox to
\@tempdima{\hfill\GRAPHIC{#5}{#4}{#1}{#2}{#3}\hfill}\\%
                  \noindent\hbox to
\@tempdima{\parbox[b]{\@tempdima}{\QCBOptA}}%
               }%
               \wd1=\@tempdima
            \else
               \textwidth=\wd0
               \setbox1 =\vbox{%
                 \noindent\hbox to
\wd0{\hfill\GRAPHIC{#5}{#4}{#1}{#2}{#3}\hfill}\\%
                 \noindent\hbox{\QCBOptA}%
               }%
               \wd1=\wd0
            \fi
         \else
            \ifdim\wd0>0pt
              \hsize=\@tempdima
              \setbox1 =\vbox{%
                \unskip\GRAPHIC{#5}{#4}{#1}{#2}{0pt}%
                \break
                \unskip\hbox to \@tempdima{\hfill \QCBOptA\hfill}%
              }%
              \wd1=\@tempdima
           \else
              \hsize=\@tempdima
              \setbox1 =\vbox{%
                \unskip\GRAPHIC{#5}{#4}{#1}{#2}{0pt}%
              }%
              \wd1=\@tempdima
           \fi
         \fi
         \@tempdimb=\ht1
         \advance\@tempdimb by \dp1
         \advance\@tempdimb by -#2%
         \advance\@tempdimb by #3%
         \leavevmode
         \raise -\@tempdimb \hbox{\box1}%
      \fi
      \egroup%
}%
\def\DFRAME#1#2#3#4#5{%
 \begin{center}
     \let\QCTOptA\empty
     \let\QCTOptB\empty
     \let\QCBOptA\empty
     \let\QCBOptB\empty
     \ifOverFrame
        #5\QCTOptA\par
     \fi
     \GRAPHIC{#4}{#3}{#1}{#2}{\z@}
     \ifUnderFrame
        \par #5\QCBOptA
     \fi
 \end{center}%
 }%
\def\FFRAME#1#2#3#4#5#6#7{%
 \begin{figure}[#1]%
  \let\QCTOptA\empty
  \let\QCTOptB\empty
  \let\QCBOptA\empty
  \let\QCBOptB\empty
  \ifOverFrame
    #4
    \ifx\QCTOptA\empty
    \else
      \ifx\QCTOptB\empty
        \caption{\QCTOptA}%
      \else
        \caption[\QCTOptB]{\QCTOptA}%
      \fi
    \fi
    \ifUnderFrame\else
      \label{#5}%
    \fi
  \else
    \UnderFrametrue%
  \fi
  \begin{center}\GRAPHIC{#7}{#6}{#2}{#3}{\z@}\end{center}%
  \ifUnderFrame
    #4
    \ifx\QCBOptA\empty
      \caption{}%
    \else
      \ifx\QCBOptB\empty
        \caption{\QCBOptA}%
      \else
        \caption[\QCBOptB]{\QCBOptA}%
      \fi
    \fi
    \label{#5}%
  \fi
  \end{figure}%
 }%
\newcount\dispkind%
\def\FRAME#1#2#3#4#5#6#7#8{%
 \ifnum\draft=\@ne
   \wasdrafttrue
 \else
   \wasdraftfalse%
 \fi
 \def\LaTeXparams{}%
 \dispkind=\z@
 \def\LaTeXparams{}%
 \doFRAMEparams{#1}%
 \ifnum\dispkind=\z@\IFRAME{#2}{#3}{#4}{#7}{#8}{#5}\else
  \ifnum\dispkind=\@ne\DFRAME{#2}{#3}{#7}{#8}{#5}\else
   \ifnum\dispkind=\tw@
    \edef\@tempa{\noexpand\FFRAME{\LaTeXparams}}%
    \@tempa{#2}{#3}{#5}{#6}{#7}{#8}%
    \fi
   \fi
  \fi
  \ifwasdraft\draft=1\else\draft=0\fi{}%
 }%
\def\TEXUX#1{"texux"}

\long\def\QQQ#1#2{%
     \long\expandafter\def\csname#1\endcsname{#2}}%
\@ifundefined{QTP}{\def\QTP#1{}}{}
\@ifundefined{Qlb}{}{}
\@ifundefined{Qlt}{}{}
\long\def\QQA#1#2{}%
\def\QTR#1#2{{\csname#1\endcsname #2}}%(gp) Is this the best?
\def\EXPAND#1[#2]#3{}%
\def\NOEXPAND#1[#2]#3{}%
\def\LaTeXparent#1{}%
\def\ChildStyles#1{}%
\def\ChildDefaults#1{}%
\def\QTagDef#1#2#3{}%
\def\QQfnmark#1{\footnotemark}

\@ifundefined{INDEX}{\def\INDEX#1#2{}{}}{}%
\@ifundefined{SUBINDEX}{\def\SUBINDEX#1#2#3{}{}{}}{}%
\def\initial#1{\bigbreak{\raggedright\large\bf #1}\kern 2\p@
   \penalty3000}%
\@ifundefined{ZZZ}{}{\makeatletter\input gnuindex.sty\makeatother\makeindex\makeatletter}%
\@ifundefined{abstract}{%
 \def\abstract{%
  \if@twocolumn
   \section*{Abstract (Not appropriate in this style!)}%
   \else \small
   \begin{center}{\bf Abstract\vspace{-.5em}\vspace{\z@}}\end{center}%
   \quotation
   \fi
  }%
 }{%
 }%
\@ifundefined{endabstract}{\def\endabstract
  {\if@twocolumn\else\endquotation\fi}}{}%
\@ifundefined{maketitle}{\def\maketitle#1{}}{}%
\@ifundefined{affiliation}{\def\affiliation#1{}}{}%
\@ifundefined{proof}{}{}%
\@ifundefined{endproof}{}{}%
\@ifundefined{newfield}{\def\newfield#1#2{}}{}%
\@ifundefined{chapter}{\def\chapter#1{\par(Chapter head:)#1\par }%
 \newcount\c@chapter}{}%
\@ifundefined{part}{\def\part#1{\par(Part head:)#1\par }}{}%
\@ifundefined{section}{\def\section#1{\par(Section head:)#1\par }}{}%
\@ifundefined{subsection}{\def\subsection#1%
 {\par(Subsection head:)#1\par }}{}%
\@ifundefined{subsubsection}{\def\subsubsection#1%
 {\par(Subsubsection head:)#1\par }}{}%
\@ifundefined{paragraph}{\def\paragraph#1%
 {\par(Subsubsubsection head:)#1\par }}{}%
\@ifundefined{subparagraph}{\def\subparagraph#1%
 {\par(Subsubsubsubsection head:)#1\par }}{}%
\@ifundefined{therefore}{}{}%
\@ifundefined{backepsilon}{}{}%
\@ifundefined{yen}{}{}%
\@ifundefined{registered}{%
   \def\registered{\relax\ifmmode{}\r@gistered
                    \else$\m@th\r@gistered$\fi}%
 \def\r@gistered{^{\ooalign
  {\hfil\raise.07ex\hbox{$\scriptstyle\rm\text{R}$}\hfil\crcr
  \mathhexbox20D}}}}{}%
\@ifundefined{Eth}{}{}%
\@ifundefined{eth}{}{}%
\@ifundefined{Thorn}{}{}%
\@ifundefined{thorn}{}{}%
\@ifundefined{degree}{}{}%
\newdimen\theight
\def\Column{%
 \vadjust{\setbox\z@=\hbox{\scriptsize\quad\quad tcol}%
  \theight=\ht\z@\advance\theight by \dp\z@\advance\theight by \lineskip
  \kern -\theight \vbox to \theight{%
   \rightline{\rlap{\box\z@}}%
   \vss
   }%
  }%
 }%
\def\qed{%
 \ifhmode\unskip\nobreak\fi\ifmmode\ifinner\else\hskip5\p@\fi\fi
 \hbox{\hskip5\p@\vrule width4\p@ height6\p@ depth1.5\p@\hskip\p@}%
 }%
\def\miss{\hbox{\vrule height2\p@ width 2\p@ depth\z@}}%
\def\tcol#1{{\baselineskip=6\p@ \vcenter{#1}} \Column}  %
\def\newfmtname{LaTeX2e}
\def\chkcompat{%
   \if@compatibility
   \else
     \usepackage{latexsym}
   \fi
}

\ifx\fmtname\newfmtname
  \DeclareOldFontCommand{\rm}{\normalfont\rmfamily}{\mathrm}
  \DeclareOldFontCommand{\sf}{\normalfont\sffamily}{\mathsf}
  \DeclareOldFontCommand{\tt}{\normalfont\ttfamily}{\mathtt}
  \DeclareOldFontCommand{\bf}{\normalfont\bfseries}{\mathbf}
  \DeclareOldFontCommand{\it}{\normalfont\itshape}{\mathit}
  \DeclareOldFontCommand{\sl}{\normalfont\slshape}{\@nomath\sl}
  \DeclareOldFontCommand{\sc}{\normalfont\scshape}{\@nomath\sc}
  \chkcompat
\fi

\def\alpha{\Greekmath 010B }%
\def\beta{\Greekmath 010C }%
\def\gamma{\Greekmath 010D }%
\def\delta{\Greekmath 010E }%
\def\epsilon{\Greekmath 010F }%
\def\zeta{\Greekmath 0110 }%
\def\eta{\Greekmath 0111 }%
\def\theta{\Greekmath 0112 }%
\def\iota{\Greekmath 0113 }%
\def\kappa{\Greekmath 0114 }%
\def\lambda{\Greekmath 0115 }%
\def\mu{\Greekmath 0116 }%
\def\nu{\Greekmath 0117 }%
\def\xi{\Greekmath 0118 }%
\def\pi{\Greekmath 0119 }%
\def\rho{\Greekmath 011A }%
\def\sigma{\Greekmath 011B }%
\def\tau{\Greekmath 011C }%
\def\upsilon{\Greekmath 011D }%
\def\phi{\Greekmath 011E }%
\def\chi{\Greekmath 011F }%
\def\psi{\Greekmath 0120 }%
\def\omega{\Greekmath 0121 }%
\def\varepsilon{\Greekmath 0122 }%
\def\vartheta{\Greekmath 0123 }%
\def\varpi{\Greekmath 0124 }%
\def\varrho{\Greekmath 0125 }%
\def\varsigma{\Greekmath 0126 }%
\def\varphi{\Greekmath 0127 }%

\def\nabla{\Greekmath 0272 }

\def\Greekmath#1#2#3#4{%
    \if@compatibility
        \ifnum\mathgroup=\symbold
           \mathchoice{\mbox{\boldmath$\displaystyle\mathchar"#1#2#3#4$}}%
                      {\mbox{\boldmath$\textstyle\mathchar"#1#2#3#4$}}%
                      {\mbox{\boldmath$\scriptstyle\mathchar"#1#2#3#4$}}%
                     
{\mbox{\boldmath$\scriptscriptstyle\mathchar"#1#2#3#4$}}%
        \else
           \mathchar"#1#2#3#4%
        \fi
    \else
        \ifnum\mathgroup=5 % For 2e
           \mathchoice{\mbox{\boldmath$\displaystyle\mathchar"#1#2#3#4$}}%
                      {\mbox{\boldmath$\textstyle\mathchar"#1#2#3#4$}}%
                      {\mbox{\boldmath$\scriptstyle\mathchar"#1#2#3#4$}}%
                     
{\mbox{\boldmath$\scriptscriptstyle\mathchar"#1#2#3#4$}}%
        \else
           \mathchar"#1#2#3#4%
        \fi
          \fi}

\newif\ifGreekBold  \GreekBoldfalse
\let\SAVEPBF=\pbf
\def\pbf{\GreekBoldtrue\SAVEPBF}%

\@ifundefined{theorem}{}{}
\@ifundefined{lemma}{}{}
\@ifundefined{corollary}{}{}
\@ifundefined{conjecture}{}{}
\@ifundefined{proposition}{}{}

\@ifundefined{axiom}{}{}
\@ifundefined{remark}{}{}
\@ifundefined{example}{}{}
\@ifundefined{exercise}{}{}
\@ifundefined{definition}{}{}

\@ifundefined{mathletters}{%
  \newcounter{equationnumber}
  \def\mathletters{%
     \addtocounter{equation}{1}
     \edef\@currentlabel{\theequation}%
     \setcounter{equationnumber}{\c@equation}
     \setcounter{equation}{0}%
     \edef\theequation{\@currentlabel\noexpand\alph{equation}}%
  }
  
}{}

\@ifundefined{BibTeX}{%
    \def\BibTeX{{\rm B\kern-.05em{\sc i\kern-.025em b}\kern-.08em
                 T\kern-.1667em\lower.7ex\hbox{E}\kern-.125emX}}}{}%
\@ifundefined{AmS}%
    {\def\AmS{{\protect\usefont{OMS}{cmsy}{m}{n}%
                A\kern-.1667em\lower.5ex\hbox{M}\kern-.125emS}}}{}%
\@ifundefined{AmSTeX}{}{}%
\ifx\ds@amstex\relax
\makeatother\endinput% 2.09 compatability
\else
   \@ifpackageloaded{amstex}%
      {\message{amstex already loaded}\makeatother\endinput}
      {}
\fi
\let\DOTSI\relax
\def\RIfM@{\relax\ifmmode}%
\def\FN@{\futurelet\next}%
\newcount\intno@
\def\iint{\DOTSI\intno@\tw@\FN@\ints@}%
\def\iiint{\DOTSI\intno@\thr@@\FN@\ints@}%
\def\iiiint{\DOTSI\intno@4 \FN@\ints@}%
\def\idotsint{\DOTSI\intno@\z@\FN@\ints@}%
\def\ints@{\findlimits@\ints@@}%
\newif\iflimtoken@
\newif\iflimits@
\def\findlimits@{\limtoken@true\ifx\next\limits\limits@true
 \else\ifx\next\nolimits\limits@false\else
 \limtoken@false\ifx\ilimits@\nolimits\limits@false\else
 \ifinner\limits@false\else\limits@true\fi\fi\fi\fi}%
\def\multint@{\int\ifnum\intno@=\z@\intdots@                          %1
 \else\intkern@\fi                                                    %2
 \ifnum\intno@>\tw@\int\intkern@\fi                                   %3
 \ifnum\intno@>\thr@@\int\intkern@\fi                                 %4
 \int}%                                                               %5
\def\multintlimits@{\intop\ifnum\intno@=\z@\intdots@\else\intkern@\fi
 \ifnum\intno@>\tw@\intop\intkern@\fi
 \ifnum\intno@>\thr@@\intop\intkern@\fi\intop}%
\def\intic@{%
    \mathchoice{\hskip.5em}{\hskip.4em}{\hskip.4em}{\hskip.4em}}%
\def\negintic@{\mathchoice
 {\hskip-.5em}{\hskip-.4em}{\hskip-.4em}{\hskip-.4em}}%
\def\ints@@{\iflimtoken@                                              %1
 \def\ints@@@{\iflimits@\negintic@
   \mathop{\intic@\multintlimits@}\limits                             %2
  \else\multint@\nolimits\fi                                          %3
  \eat@}%                                                             %4
 \else                                                                %5
 \def\ints@@@{\iflimits@\negintic@
  \mathop{\intic@\multintlimits@}\limits\else
  \multint@\nolimits\fi}\fi\ints@@@}%
\def\intkern@{\mathchoice{\!\!\!}{\!\!}{\!\!}{\!\!}}%
\def\plaincdots@{\mathinner{\cdotp\cdotp\cdotp}}%
\def\intdots@{\mathchoice{\plaincdots@}%
 {{\cdotp}\mkern1.5mu{\cdotp}\mkern1.5mu{\cdotp}}%
 {{\cdotp}\mkern1mu{\cdotp}\mkern1mu{\cdotp}}%
 {{\cdotp}\mkern1mu{\cdotp}\mkern1mu{\cdotp}}}%
\def\RIfM@{\relax\protect\ifmmode}
\def\text{\RIfM@\expandafter\text@\else\expandafter\mbox\fi}
\let\nfss@text\text
\def\text@#1{\mathchoice
   {\textdef@\displaystyle\f@size{#1}}%
   {\textdef@\textstyle\tf@size{\firstchoice@false #1}}%
   {\textdef@\textstyle\sf@size{\firstchoice@false #1}}%
   {\textdef@\textstyle \ssf@size{\firstchoice@false #1}}%
   \glb@settings}

\def\textdef@#1#2#3{\hbox{{%
                    \everymath{#1}%
                    \let\f@size#2\selectfont
                    #3}}}
\newif\iffirstchoice@
\firstchoice@true
\def\Let@{\relax\iffalse{\fi\let\\=\cr\iffalse}\fi}%
\def\vspace@{\def\vspace##1{\crcr\noalign{\vskip##1\relax}}}%
\def\multilimits@{\bgroup\vspace@\Let@
 \baselineskip\fontdimen10 \scriptfont\tw@
 \advance\baselineskip\fontdimen12 \scriptfont\tw@
 \lineskip\thr@@\fontdimen8 \scriptfont\thr@@
 \lineskiplimit\lineskip
 \vbox\bgroup\ialign\bgroup\hfil$\m@th\scriptstyle{##}$\hfil\crcr}%
\def\Sb{_\multilimits@}%
\def\endSb{\crcr\egroup\egroup\egroup}%
\def\Sp{^\multilimits@}%

\newdimen\ex@
\def\rightarrowfill@#1{$#1\m@th\mathord-\mkern-6mu\cleaders
 \hbox{$#1\mkern-2mu\mathord-\mkern-2mu$}\hfill
 \mkern-6mu\mathord\rightarrow$}%
\def\leftarrowfill@#1{$#1\m@th\mathord\leftarrow\mkern-6mu\cleaders
 \hbox{$#1\mkern-2mu\mathord-\mkern-2mu$}\hfill\mkern-6mu\mathord-$}%
\def\leftrightarrowfill@#1{$#1\m@th\mathord\leftarrow
\mkern-6mu\cleaders
 \hbox{$#1\mkern-2mu\mathord-\mkern-2mu$}\hfill
 \mkern-6mu\mathord\rightarrow$}%
\def\overrightarrow{\mathpalette\overrightarrow@}%
\def\overrightarrow@#1#2{\vbox{\ialign{##\crcr\rightarrowfill@#1\crcr
 \noalign{\kern-\ex@\nointerlineskip}$\m@th\hfil#1#2\hfil$\crcr}}}%

\def\overleftarrow{\mathpalette\overleftarrow@}%
\def\overleftarrow@#1#2{\vbox{\ialign{##\crcr\leftarrowfill@#1\crcr
 \noalign{\kern-\ex@\nointerlineskip}$\m@th\hfil#1#2\hfil$\crcr}}}%
\def\overleftrightarrow{\mathpalette\overleftrightarrow@}%
\def\overleftrightarrow@#1#2{\vbox{\ialign{##\crcr
   \leftrightarrowfill@#1\crcr
 \noalign{\kern-\ex@\nointerlineskip}$\m@th\hfil#1#2\hfil$\crcr}}}%
\def\underrightarrow{\mathpalette\underrightarrow@}%
\def\underrightarrow@#1#2{\vtop{\ialign{##\crcr$\m@th\hfil#1#2\hfil
  $\crcr\noalign{\nointerlineskip}\rightarrowfill@#1\crcr}}}%

\def\underleftarrow{\mathpalette\underleftarrow@}%
\def\underleftarrow@#1#2{\vtop{\ialign{##\crcr$\m@th\hfil#1#2\hfil
  $\crcr\noalign{\nointerlineskip}\leftarrowfill@#1\crcr}}}%
\def\underleftrightarrow{\mathpalette\underleftrightarrow@}%
\def\underleftrightarrow@#1#2{\vtop{\ialign{##\crcr$\m@th
  \hfil#1#2\hfil$\crcr
 \noalign{\nointerlineskip}\leftrightarrowfill@#1\crcr}}}%
\def\qopnamewl@#1{\mathop{\operator@font#1}\nlimits@}
\let\nlimits@\displaylimits
\def\setboxz@h{\setbox\z@\hbox}

\def\varlim@#1#2{\mathop{\vtop{\ialign{##\crcr
 \hfil$#1\m@th\operator@font lim$\hfil\crcr
 \noalign{\nointerlineskip}#2#1\crcr
 \noalign{\nointerlineskip\kern-\ex@}\crcr}}}}

 \def\rightarrowfill@#1{\m@th\setboxz@h{$#1-$}\ht\z@\z@
  $#1\copy\z@\mkern-6mu\cleaders
  \hbox{$#1\mkern-2mu\box\z@\mkern-2mu$}\hfill
  \mkern-6mu\mathord\rightarrow$}
\def\leftarrowfill@#1{\m@th\setboxz@h{$#1-$}\ht\z@\z@
  $#1\mathord\leftarrow\mkern-6mu\cleaders
  \hbox{$#1\mkern-2mu\copy\z@\mkern-2mu$}\hfill
  \mkern-6mu\box\z@$}

\def\projlim{\qopnamewl@{proj\,lim}}
\def\injlim{\qopnamewl@{inj\,lim}}
\def\varinjlim{\mathpalette\varlim@\rightarrowfill@}
\def\varprojlim{\mathpalette\varlim@\leftarrowfill@}
\def\varliminf{\mathpalette\varliminf@{}}
\def\varliminf@#1{\mathop{\underline{\vrule\@depth.2\ex@\@width\z@
   \hbox{$#1\m@th\operator@font lim$}}}}
\def\varlimsup{\mathpalette\varlimsup@{}}
\def\varlimsup@#1{\mathop{\overline
  {\hbox{$#1\m@th\operator@font lim$}}}}

\def\dfrac#1#2{{\displaystyle {#1 \over #2}}}%
\begingroup \catcode `|=0 \catcode `[= 1
\catcode`]=2 \catcode `\{=12 \catcode `\}=12
\catcode`\\=12
|gdef|@alignverbatim#1\end{align}[#1|end[align]]
|gdef|@salignverbatim#1\end{align*}[#1|end[align*]]

|gdef|@alignatverbatim#1\end{alignat}[#1|end[alignat]]
|gdef|@salignatverbatim#1\end{alignat*}[#1|end[alignat*]]

|gdef|@xalignatverbatim#1\end{xalignat}[#1|end[xalignat]]
|gdef|@sxalignatverbatim#1\end{xalignat*}[#1|end[xalignat*]]

|gdef|@gatherverbatim#1\end{gather}[#1|end[gather]]
|gdef|@sgatherverbatim#1\end{gather*}[#1|end[gather*]]

|gdef|@gatherverbatim#1\end{gather}[#1|end[gather]]
|gdef|@sgatherverbatim#1\end{gather*}[#1|end[gather*]]

|gdef|@multilineverbatim#1\end{multiline}[#1|end[multiline]]
|gdef|@smultilineverbatim#1\end{multiline*}[#1|end[multiline*]]

|gdef|@arraxverbatim#1\end{arrax}[#1|end[arrax]]
|gdef|@sarraxverbatim#1\end{arrax*}[#1|end[arrax*]]

|gdef|@tabulaxverbatim#1\end{tabulax}[#1|end[tabulax]]
|gdef|@stabulaxverbatim#1\end{tabulax*}[#1|end[tabulax*]]

|endgroup

\def\align{\@verbatim \frenchspacing\@vobeyspaces \@alignverbatim
You are using the "align" environment in a style in which it is not
defined.}

\@namedef{align*}{\@verbatim\@salignverbatim
You are using the "align*" environment in a style in which it is not
defined.}
\expandafter\let\csname endalign*\endcsname =\endtrivlist

\def\alignat{\@verbatim \frenchspacing\@vobeyspaces \@alignatverbatim
You are using the "alignat" environment in a style in which it is not
defined.}

\@namedef{alignat*}{\@verbatim\@salignatverbatim
You are using the "alignat*" environment in a style in which it is not
defined.}
\expandafter\let\csname endalignat*\endcsname =\endtrivlist

\def\xalignat{\@verbatim \frenchspacing\@vobeyspaces \@xalignatverbatim
You are using the "xalignat" environment in a style in which it is not
defined.}

\@namedef{xalignat*}{\@verbatim\@sxalignatverbatim
You are using the "xalignat*" environment in a style in which it is not
defined.}
\expandafter\let\csname endxalignat*\endcsname =\endtrivlist

\def\gather{\@verbatim \frenchspacing\@vobeyspaces \@gatherverbatim
You are using the "gather" environment in a style in which it is not
defined.}

\@namedef{gather*}{\@verbatim\@sgatherverbatim
You are using the "gather*" environment in a style in which it is not
defined.}
\expandafter\let\csname endgather*\endcsname =\endtrivlist

\def\multiline{\@verbatim \frenchspacing\@vobeyspaces \@multilineverbatim
You are using the "multiline" environment in a style in which it is not
defined.}

\@namedef{multiline*}{\@verbatim\@smultilineverbatim
You are using the "multiline*" environment in a style in which it is not
defined.}
\expandafter\let\csname endmultiline*\endcsname =\endtrivlist

\def\arrax{\@verbatim \frenchspacing\@vobeyspaces \@arraxverbatim
You are using a type of "array" construct that is only allowed in
AmS-LaTeX.}

\def\tabulax{\@verbatim \frenchspacing\@vobeyspaces \@tabulaxverbatim
You are using a type of "tabular" construct that is only allowed in
AmS-LaTeX.}

\@namedef{arrax*}{\@verbatim\@sarraxverbatim
You are using a type of "array*" construct that is only allowed in
AmS-LaTeX.}
\expandafter\let\csname endarrax*\endcsname =\endtrivlist

\@namedef{tabulax*}{\@verbatim\@stabulaxverbatim
You are using a type of "tabular*" construct that is only allowed in
AmS-LaTeX.}
\expandafter\let\csname endtabulax*\endcsname =\endtrivlist

\def\@@eqncr{\let\@tempa\relax
    \ifcase\@eqcnt \def\@tempa{& & &}\or \def\@tempa{& &}%
      \else \def\@tempa{&}\fi
     \@tempa
     \if@eqnsw
        \iftag@
           \@taggnum
        \else
           \@eqnnum\stepcounter{equation}%
        \fi
     \fi
     \global\tag@false
     \global\@eqnswtrue
     \global\@eqcnt\z@\cr}

 \def\endequation{%
     \ifmmode\ifinner % FLEQN hack
      \iftag@
        \addtocounter{equation}{-1} % undo the increment made in the begin
part
        $\hfil
           \displaywidth\linewidth\@taggnum\egroup \endtrivlist
        \global\tag@false
        \global\@ignoretrue
      \else
        $\hfil
           \displaywidth\linewidth\@eqnnum\egroup \endtrivlist
        \global\tag@false
        \global\@ignoretrue
      \fi
     \else
      \iftag@
        \addtocounter{equation}{-1} % undo the increment made in the begin
part
        \eqno \hbox{\@taggnum}
        \global\tag@false%
        $$\global\@ignoretrue
      \else
        \eqno \hbox{\@eqnnum}% $$ BRACE MATCHING HACK
        $$\global\@ignoretrue
      \fi
     \fi\fi
 }

 \newif\iftag@ \tag@false

 \def\tag{\@ifnextchar*{\@tagstar}{\@tag}}
 \def\@tag#1{%
     \global\tag@true
     \global\def\@taggnum{(#1)}}
 \def\@tagstar*#1{%
     \global\tag@true
     \global\def\@taggnum{#1}%
}

\makeatother
\QQQ{Language}{American English}
\def\ba{\begin{array}}
\def\ea{\end{array}}
\def\beq{\begin{equation}}
\def\eeq{\end{equation}}
\def\bea{\begin{eqnarray}}
\def\eea{\end{eqnarray}}

\begin{document}

\renewcommand{\thefootnote}{\fnsymbol{footnote}} \newpage
\pagestyle{empty} \setcounter{page}{0}

%%%%%%%%%%%%%%%%%%%%%%%%%%%%%%%%%%%%%%%%%%%%%%%%%%%%%%%%%%%%%%%
%%%%%%%%%%%%%%%%%%%% LOGO ENSLAPP - DEBUT  %%%%%%%%%%%%%%%%%%%%
%%%%%%%%%%%%%%%%%%%%%%%%%%%%%%%%%%%%%%%%%%%%%%%%%%%%%%%%%%%%%%%
\newcommand{\norm}[1]{{\protect\normalsize{#1}}}
\newcommand{\LAP}
{{\small E}\norm{N}{\large S}{\Large L}{\large A}\norm{P}{\small P}}
\newcommand{\sLAP}{{\scriptsize E}{\footnotesize{N}}{\small S}{\norm L}$
${\small A}{\footnotesize{P}}{\scriptsize P}}
\def\logolapin{
  \raisebox{-1.2cm}{\epsfbox{enslapp.ps}}}
\def\logolight{{\bf{{\large E}{\Large N}{\LARGE S}{\huge L}{\LARGE
        A}{\Large P}{\large P}} }}
\def\logoenslapp{\logolight}
%%%%%%%%%%%%%%%%%%%%%%%%%%%%%%%%%%%%%%%%%%%%%%%%%%%%%%%%%%%%%%%
%%%% NB!!! POUR GAGNER DU TEMPS OU ENVOYER A L'EXTERIEUR, %%%%%
%%%% ZAPPER LE LAPIN, I.E. COMMENTER LA LIGNE DE COMMANDE %%%%%
%%%% QUI CONTIENT \logolapin CI-DESSOUS                   %%%%%
%
%\def\logoenslapp{\logolapin}\input epsf
%
%%%%%%%%%%%%%%%%%%%%%%%%%%%%%%%%%%%%%%%%%%%%%%%%%%%%%%%%%%%%%%%
%
\hbox to \hsize{
\hss
\begin{minipage}{5.2cm}
  \begin{center}
    {\bf Groupe d'Annecy\\ \ \\
      Laboratoire d'Annecy-le-Vieux de Physique des Particules}
  \end{center}
\end{minipage}
\hfill
\logoenslapp
\hfill
\begin{minipage}{4.2cm}
  \begin{center}
    {\bf Groupe de Lyon\\ \ \\
      Ecole Normale Supérieure de Lyon}
  \end{center}
\end{minipage}
\hss}

\vspace {.3cm}
\centerline{\rule{12cm}{.42mm}}
%%%%%%%%%%%%%%%%%%%%%%%%%%%%%%%%%%%%%%%%%%%%%%%%%%%%%%%%%%%%%%%
%%%%%%%%%%%%%%%%%%%%% LOGO ENSLAPP  - FIN %%%%%%%%%%%%%%%%%%%%%
%%%%%%%%%%%%%%%%%%%%%%%%%%%%%%%%%%%%%%%%%%%%%%%%%%%%%%%%%%%%%%%

\vspace{20mm}

\begin{center}
{\Large {\bf SOME REMARKS ON TOPOLOGICAL 4d-GRAVITY}}%
\\[1cm]

\vspace{10mm}

{\large Frank Thuillier}

{\em Laboratoire de Physique Th\'eorique }{\small E}{{\normalsize
{N}}} {\large S}{\Large L}{\large A}{{\normalsize {P}}}{\small P}
\footnote{URA 14-36 du CNRS, associ\'ee \`a l'Ecole Normale
Sup\'erieure de Lyon et \`a l'Universit\'e de Savoie} \\ {\em LAPP,
Chemin de Bellevue BP 110, F-74941 Annecy-le-Vieux Cedex, France}
\end{center}

\vspace{20mm}

\centerline{ {\bf Abstract}}

\vspace{5mm}

\indent We show that the method of S. Wu to study topological 4d-gravity
can be understood within a now standard method designed to produce
equivariant cohomology classes. Next, this general framework is
applied to produce some observables of the topological 4d-gravity.

\vfill
\rightline{hep-th/9707084}
\rightline{\LAP-A-648/97} \rightline{ 1997}

\newpage
\pagestyle{plain} \renewcommand{\thefootnote}{\arabic{footnote}} \newpage

\pagestyle{plain} \pagenumbering{arabic}

\section{Introduction.}

Since its appearance in 1988 in a famous article of E. Witten
\cite{W88}, Topological Field Theories have played an important role
in theoretical physics as well as in mathematics. Actually, the 1988
article gave a prototype of Topological Field Theories of
cohomological type. Witten has recognized that these Cohomological
Field Theories are related to Equivariant Cohomology and more
precisely to the so-called Cartan model of Equivariant Cohomology.

Although Cohomological Field Theories can be described
independently of which model is used for Equivariant Cohomology,
the construction by J. Kalkman (\cite{K93}) of the so-called
intermediate model (\cite{STW94}) is of considerable technical
help. In \cite{STW94}, Topological Yang-Mills
(\cite{W88,BS88,AJ90}) and Topological 2d Gravity (\cite{BS91})
where studied from this point of view. In \cite{BT97}, new
representatives of the Thom class of a vector bundle where produced
using this general framework.

S. Wu \cite{Wu93} explained the role of the universal bundle in 4d
Gravity, and exhibited some observables of the corresponding
topological model. We shall explained here how his method can be
deduced from the general approach of \cite{STW94} and which
observables are obtained.

\section{From the Intermediate to the Weil Model of equivariant
cohomology.\label{sec2}}

In \cite{STW94} it was explained how one can generate representatives
of equivariant cohomology classes using an idea of \cite{BGV91} which
benefits from J. Kalkman's construction (\cite{K93}) as follows: let
us assume that ${\cal M}$ is a smooth manifold with smooth ${\cal
G}$-action for some connected Lie group ${\cal G}$ (with Lie algebra
Lie${\cal G}$). Let $d_{{\cal M}}$, $i_{{\cal M}}$, $l_{{\cal M}}$ be
the standard exterior derivative, inner product and Lie derivative on
${\cal M}$. The action of ${\cal G}$ induces an action of Lie${\cal
G}$, and to any $\lambda
\in $ Lie${\cal G}$, there corresponds a so-called fundamental vector field
$\lambda _{{\cal M}}$ on ${\cal M}$. The space of forms on ${\cal
M}$ is denoted by $\Omega ({\cal M})$, and its basic elements are
those annihilated both by $i_{{\cal M}}(\lambda )$ and $l_{{\cal
M}}(\lambda )$, for any $ \lambda \in $Lie${\cal G}$. We recall
that $l_{{\cal M}}=\left[ d_{{\cal M} },i_{{\cal M}}\right]_{+} $.

The Weil algebra $({\cal W}({\cal G})$,$d_{{\cal W}},$ $i_{{\cal
W}}$, $l_{ {\cal W}})$ of ${\cal G}$, is the graded differential
algebra generated by the "connection $\omega $" and its "curvature
$\Omega $":

\begin{center}
\begin{eqnarray}
d_{{\cal W}}\omega &=&\Omega -\dfrac 12\left[ \omega ,\omega \right] \\
d_{{\cal W}}\Omega &=&-\left[ \omega ,\Omega \right] \\
i_{{\cal W}}(\lambda )\omega &=&\lambda \\
i_{{\cal W}}(\lambda )\Omega &=&0 \\
l_{{\cal W}}(\lambda )\omega &=&-\left[ \lambda ,\omega \right] \\
i_{{\cal W}}(\lambda )\Omega &=&-\left[ \lambda ,\Omega \right]
\end{eqnarray}
\end{center}

\noindent for any $\lambda \in $Lie${\cal G}$.

Then, the equivariant cohomology for the action of ${\cal G}$ on
${\cal M}$ is the basic cohomology of the graded differential
algebra $({\cal W}({\cal G})\otimes \Omega ({\cal M}),d_{{\cal
W}}+d_{{\cal M}},i_{{\cal W}}+i_{{\cal M}},l_{{\cal W}}+l_{{\cal
M}})$. It generates the so-called Weil model of equivariant
cohomology.

Now let us consider another Lie group $H$ such that ${\cal M}$ is
the base space of some principal $H$-bundle ${\cal P}({\cal M},H)$
opn which the action of ${\cal G}$ can be lifted. This bundle is
also equipped with standard differential operations: $d_{{\cal
P}}$, $i_{ {\cal P}}$, $l_{{\cal P}}$. Then, some equivariant
cohomology classes can be represented as follows: consider a ${\cal
G}$-invariant $H$-connection $\Gamma$ on ${\cal P}$. Extend $\Gamma
$ to $ {\cal W}({\cal G})\otimes\Omega ({\cal M})$, still denoting
it $\Gamma $. Since $\Gamma $ does not depend on $\omega $, it
fulfills :

\begin{eqnarray}
i_{{\cal W}}(\lambda )\Gamma &=&0 \\
(l_{{\cal W}}+l_{{\cal P}})(\lambda )\Gamma &=&0
\end{eqnarray}

\noindent for any $\lambda \in $Lie${\cal G}$. That expresses the basicity
of $\Gamma $ in the so-called Intermediate model of equivariant cohomology.
In this model, the exterior derivative reads :

\begin{equation}
D_{int}^{}=d_{{\cal W}}+d_{{\cal P}}+l_{{\cal P}}(\omega )-i_{{\cal
P}}(\Omega )
\end{equation}

\noindent so that :

\begin{equation}
D_{int}^{}\Gamma =d_{{\cal P}}\Gamma -i_{{\cal P}}(\Omega )\Gamma
\end{equation}

\noindent and the Equivariant curvature of $\Gamma $
in the intermediate model reads  :

\begin{equation}
R_{int}^{eq}(\Gamma ,\omega ,\Omega )=D_{int}^{}\Gamma +\dfrac 12\left[
\Gamma ,\Gamma \right]
\end{equation}

\noindent It satisfies :

\begin{eqnarray}
D_{int}^{}R_{int}^{eq} &=&\left[ R_{int}^{eq},\Gamma \right] \\
i_{{\cal W}}(\lambda )R_{int}^{eq} &=&0 \\
(l_{{\cal W}}+l_{{\cal P}})(\lambda )R_{int}^{eq} &=&0
\end{eqnarray}

\noindent The $H$-fibration is eliminated by considering symmetric
$H$-invariant polynomials $I_{int}^{eq}=I(R_{int}^{eq})$.

To go to the more usual Weil Model, we use the Kalkman differential
algebra isomorphism $\exp \left\{ i_{{\cal P}}(\omega )\right\} $,
thus obtaining:

\begin{eqnarray}
(d_{{\cal W}}+d_{{\cal P}})I_W^{eq} &=&0 \\
(i_{{\cal W}}+i_{{\cal P}})(\lambda )I_W^{eq} &=&0 \\
(l_{{\cal W}}+l_{{\cal P}})(\lambda )I_W^{eq} &=&0
\end{eqnarray}

\noindent where $I_W^{eq}$=$\exp \left\{ i_{{\cal P}}(\omega )\right\}
I_{int}^{eq}$. Now since the $H$-fibration has disappeared,
$I_W^{eq}$ lies in ${\cal W}({\cal G})\otimes \Omega ({\cal M})$.
Under the assumption that ${\cal M}$ is a principal ${\cal
M}$-bundle over ${\cal M}$/${\cal G}$, we can replace $\omega $ and
$\Omega $ by a ${\cal G}$-connection $\theta $ and its curvature
$\Theta $ on ${\cal M}$. Cartan's theorem 3 guarantees that our new
representative gives a representative of the same equivariant
cohomology class (\cite{C50}, \cite{STW94}). Still denoting this
representative by $I_W^{eq}$, we verify that:

\begin{eqnarray}
d_{{\cal M}}I_W^{eq} &=&0 \\
i_{{\cal M}}(\lambda )I_W^{eq} &=&0 \\
l_{{\cal M}}(\lambda )I_W^{eq} &=&0
\end{eqnarray}

\noindent Now, we are ready to use this method in topological 4d-gravity.

\section{Wu's construction (\cite{Wu93}) in Topological 4d Gravity.}

Let $\Sigma $ be a 4d smooth manifold. The fundamental objects in
$Gr_4^{top} $ are the metrics of $\Sigma $, and the generators of
the Weil algebra of $Diff_{0}(\Sigma )$ the connected component of
the diffeomorphism group of $\Sigma $. The structure equations then
read :

\begin{eqnarray}
s^{top}g &=&\Psi +L^{top}(\omega )g \\
s^{top}\Psi &=&-L^{top}(\Omega )g+L^{top}(\omega )\Psi \\
s^{top}\omega &=&\Omega -\dfrac 12\left[ \omega ,\omega \right] \\
s^{top}\Omega &=&-\left[ \omega ,\Omega \right]
\end{eqnarray}

\noindent Let us note that
the form of these structure equations is universal (i.e. independent
of the model we choose). Now let us apply the precepts of the previous
section. The group of diffeomorphisms of $\Sigma $ plays the role of
the gauge group ${\cal G}$ over $Met(\Sigma )$. The $H$-fibration is
obtained by considering the frame bundle over $\Sigma $, $F(\Sigma )$
\footnote[1]{Note that $F(\Sigma )$ is the principal bundle associated
to the tangent vector bundle $T\Sigma$ of $\Sigma$}, and our final
principal $GL(4,{\bf R})$-bundle ${\cal P}$ is just $Met(\Sigma
)\times F(\Sigma )$. The $Diff(\Sigma )$-invariant $GL(4,{\bf
R})$-connection $\Gamma $ on $Met(\Sigma )\times F(\Sigma )$ is given
by:

\begin{equation}
\Gamma _{\text{\quad }\mu }^\lambda =\Gamma ^{LC}(g)_{\text{\quad }\mu
}^\lambda +\dfrac 12g^{\lambda \nu }\delta g_{\nu \mu }
\end{equation}

\noindent where $\Gamma ^{LC}(g)$ is the Levi-Civita connection of $g\in
Met(\Sigma )$, and $\delta $ is the exterior derivative on
$Met(\Sigma )$ (\cite {BS91,DV}).

This $GL(4,{\bf R})$-connection is used in the Intermediate Model.
Before going any further, let us notice that in the Weil model,
this connection reads:

\begin{equation}
\tilde{\Gamma}_{\text{\quad }\mu }^\lambda =\Gamma _{\text{\quad }\mu
}^\lambda -(i_{{\cal P}}(\Omega )\Gamma )_{\text{\quad }\mu }^\lambda
\end{equation}

\noindent which is comparable with (2.5) in \cite{Wu93}. Now, the
Intermediate curvature :

\begin{equation}
R_{int}^{eq}(\Gamma ,\omega ,\Omega )=D_{int}^{}\Gamma +\dfrac 12\left[
\Gamma ,\Gamma \right]
\end{equation}

\noindent gives the corresponding Weil curvature :

\begin{equation}
R_W^{eq}(\Gamma ,\omega ,\Omega )=\exp \left\{ i_{{\cal P}}(\omega
)\right\} R_{int}^{eq}(\Gamma ,\omega ,\Omega )=(d_{{\cal
W}}+d_{{\cal P}})\tilde{\Gamma}+\dfrac 12\left[
\tilde{\Gamma},\tilde{\Gamma}\right]
\end{equation}

\noindent which is of the form (2.6) of \cite{Wu93}.

Now, let us have a look at the observables.

\section{Some Observables for Topological 4d Gravity.}

In oder to generate observables of the theory, we first eliminate the
$GL(4,{\bf R})$-fibration. As explained in section \ref{sec2} this is
achieved by considering symmetric $GL(4,{\bf R})$-invariant
polynomials. The Euler class and the Pontrjagin classes generated by
$R_W^{eq}$ are such polynomials (\cite{KN63}). Actually, only the
first Pontrjagin class is relevant \footnote{The zeroth class is
trivially 1 while the second (and highest) class is the square of the
Euler class.}. Up to normalization factors, those two cohomology
classes are given by:

\begin{eqnarray}
E_W^{eq}&=&\dfrac{\varepsilon ^{\mu \nu \rho \sigma }}{\sqrt{{\bf
g}}}g_{\upsilon \lambda }g_{\sigma \chi }\left( R_W^{eq}\right)
_\mu ^\lambda \wedge \left( R_W^{eq}\right) _\rho ^\chi \\
P_W^{eq}&=&\left(\delta^{\mu}_{\lambda} \delta^{\rho}_{\chi} -
\delta^{\mu}_{\chi} \delta^{\rho}_{\lambda} \right)
\left( R_W^{eq}\right)
_\mu ^\lambda
\wedge \left( R_W^{eq}\right) _\rho ^\chi
\end{eqnarray}

\newpage

\noindent and decompose into five terms :

\begin{eqnarray}
E_W^{eq}&=&Q_0^4+Q_1^3+Q_2^2+Q_3^1+Q_4^0 \\
P_W^{eq}&=&G_0^4+G_1^3+G_2^2+G_3^1+G_4^0
\end{eqnarray}

\noindent where the upper index refers to the form degree on $Met(\Sigma )$
while the lower one refers to the form degree on $\Sigma $. These
expressions are to be compared with (2.9) of \cite {Wu93}. Observables
extracted from monomials $\left( E_W^{eq}\right)^m
\left( P_W^{eq}\right)^n $:

\begin{eqnarray}
\left( E_W^{eq}\right)^m\left(
P_W^{eq}\right)^n=V_0^{4(m+n)}+V_1^{4(m+n)-1}
+V_2^{4(m+n)-2}+V_3^{4(m+n)-3}+V_4^{4(m+n)-4}
\end{eqnarray}

\noindent with:

\begin{eqnarray}
V_0^{4(m+n)} &=&\left(Q_0^4\right)^m \left(G_0^4\right)^n \\
\nonumber \\[2mm]
V_1^{4(m+n)-1}&=&n\left(Q_0^4\right)^m \left(G_0^4\right)^{n-1}
G_1^3 + m\left(Q_0^4\right)^{m-1}Q_1^3 \left(G_0^4\right)^n \\
\nonumber \\[2mm]
V_2^{4(m+n)-2}&=&n\left(Q_0^4\right)^m
\left(G_0^4\right)^{n-1}G_2^2 +\frac{n(n-1)}2 \left(Q_0^4\right)^m
\left(G_0^4\right)^{n-2} \left( G_1^3\right)^2 \\
&&+mn\left(Q_0^4\right)^{m-1}Q_1^3 \left(G_0^4\right)^{n-1} G_1^3
+m\left(Q_0^4\right)^{m-1}Q_2^2 \left(G_0^4\right)^n \nonumber \\
&&+\frac{m(m-1)}2 \left( Q_0^4\right)^{m-2}Q_2^2 Q_1^3
\left(G_0^4\right)^n \nonumber \\
\nonumber \\[2mm]
V_3^{4(m+n)-3}&=&n\left(Q_0^4\right)^m\left(G_0^4\right)^{n-1}G_3^1
+\frac{n(n-1)}2 \left(Q_0^4\right)^m \left(G_0^4\right)^{n-2}G_2^2
G_1^3 \\ && +\frac{n(n-1)(n-2)}6 \left(Q_0^4\right)^m
\left(G_0^4\right)^{n-3} \left( G_1^3\right)^3 \nonumber \\
&& +mn\left(Q_0^4\right)^{m-1}Q_1^3 \left(G_0^4\right)^{n-1}G_2^2
+m\frac{n(n-1)}2\left(Q_0^4\right)^{m-1}Q_1^3
\left(G_0^4\right)^{n-2} \left( G_1^3\right)^2 \nonumber \\
&& +mn\left(Q_0^4\right)^{m-1}Q_2^2 \left(G_0^4\right)^{n-1} G_1^3
+n\frac{m(m-1)}2 \left( Q_0^4\right)^{m-2} \left( Q_1^3\right)^2
\left(G_0^4\right)^{n-1} G_1^3 \nonumber \\
&& +m\left(Q_0^4\right)^{m-1}Q_3^1 \left(G_0^4\right)^n
+\frac{m(m-1)}2 \left(Q_0^4\right)^{m-2}Q_2^2 Q_1^3
\left(G_0^4\right)^n  \nonumber \\ && +\frac{m(m-1)(m-2)}6
\left(Q_0^4\right)^{m-3} \left( Q_1^3\right)^3 \left(G_0^4\right)^n
\nonumber \\
\nonumber \\[2mm]
V_4^{4(m+n)-4} &=&n\left(Q_0^4\right)^m \left(
G_0^4\right)^{n-1}G_4^0 +\frac{n(n-1)}2 \left(Q_0^4\right)^m \left(
G_0^4\right)^{n-2}\left( \left(G_2^2\right)^2 +G_1^3G_3^1\right) \\
&& +\frac{n(n-1)(n-2)}6\left(Q_0^4\right)^m\left(G_0^4\right)^{n-3}
\left(G_1^3\right)^2 G_2^2 \nonumber \\
&&+\frac{n(n-1)(n-2)(n-3)}{24} \left(Q_0^4\right)^m \left(
G_0^4\right)^{n-4} \left( G_1^3\right)^4 \nonumber \\ &&
+mn\left(Q_0^4\right)^{m-1}Q_1^3 \left(G_0^4\right)^{n-1}G_3^1 +
m\frac{n(n-1)}2\left(Q_0^4\right)^{m-1}Q_1^3
\left(G_0^4\right)^{n-2}G_2^2G_1^3 \nonumber \\
&& +m\frac{n(n-1)(n-2)}6 \left(Q_0^4\right)^{m-1}Q_1^3
\left(G_0^4\right)^{n-3} \left( G_1^3\right)^3 \nonumber \\
&& +mn\left(Q_0^4\right)^{m-1}Q_2^2\left(G_0^4\right)^{n-1}G_2^2
+m\frac{n(n-1)}2\left(Q_0^4\right)^{m-1}Q_2^2
\left(G_0^4\right)^{n-2} \left(G_1^3\right)^2 \nonumber \\
&& +n\frac{m(m-1)}2\left(Q_0^4\right)^{m-2} \left(Q_1^3\right)^2
\left(G_0^4\right)^{n-1}G_2^2 \nonumber \\
&& +\frac{mn(m-1)(n-1)}4\left(Q_0^4\right)^{m-2}
\left(Q_1^3\right)^2\left(G_0^4\right)^{n-2} \left(G_1^3\right)^2
\nonumber \\
&& +mn\left(Q_0^4\right)^{m-1}Q_3^1 \left(G_0^4\right)^{n-1}G_1^3
+n\frac{m(m-1)}2 \left(Q_0^4\right)^{m-2}Q_2^2Q_1^3
\left(G_0^4\right)^{n-1}G_1^3 \nonumber \\
&& +n\frac{m(m-1)(m-2)}6 \left(Q_0^4\right)^{m-3}Q_1^3
\left(G_0^4\right)^{n-1}G_1^3 \nonumber \\
&& +m\left( Q_0^4\right)^{m-1}Q_4^0\left(G_0^4\right)^n
+\frac{m(m-1)}2 \left( Q_0^4\right)^{m-2}\left(
\left(Q_2^2\right)^2 +Q_1^3Q_3^1\right)\left(G_0^4\right)^n
\nonumber \\
&& +\frac{m(m-1)(m-2)}6 \left(Q_0^4\right)^{m-3}
\left(Q_1^3\right)^2 Q_2^ 2\left(G_0^4\right)^n \nonumber \\
&&+\frac{m(m-1)(m-2)(m-3)}{24} \left( Q_0^4\right)^{m-4} \left(
Q_1^3\right)^4 \left(G_0^4\right)^n \nonumber
\end{eqnarray}

\noindent Next, we replace $\omega $ and $\Omega $ by a
$Diff(\Sigma )$-connection $\theta $ and its curvature $\Theta $ on
$Met(\Sigma )$. The corresponding forms fulfill the ''descent''
equations:

\begin{eqnarray}
\delta V_p^{4n-p}+d_\Sigma V_{p-1}^{4n-p+1} &=&0 \\
{\cal I}(\lambda )V_p^{4n-p}+i_\Sigma (\lambda )V_{p+1}^{4n-p-1} &=&0
\\ {\cal L}(\lambda )V_p^{4n-p}+l_\Sigma (\lambda )V_p^{4n-p} &=&0
\end{eqnarray}

\noindent where ${\cal I}$ and ${\cal L}$ are the inner product and Lie
derivative on $Met(\Sigma )$. Finally, we integrate over cycles on $\Sigma
$
to obtain forms on $Met(\Sigma )$ only :

\begin{equation}
V^{4n-p}=\oint_{\gamma _p}V_p^{4n-p}
\end{equation}

\noindent Exactly as in the 2d gravity, only :

\begin{equation}
V^{4n-4}=\oint_\Sigma V_4^{4n-4}
\end{equation}

\noindent defines an equivariant form on $Met(\Sigma )$. This gives
observables of $Gr_4^{top}$. which are the analogs of the Mumford
invariants appearing in $Gr_2^{top}$.

An explicit expression of the Q's and the G's is given in appendix.

\section{Conclusion.}
All the work done above can be apply to higher dimensional gravity
theory. Of course this also apply to Yang-Mills topological theory.
Nevertheless, in this last case things are much simpler since the
gauge group doesn't act on the space-time manifold $\Sigma $, while in
Gravity theory the diffeomorphism group does.

\vspace{1cm}
\noindent
{\bf Acknowledgement} I would like to thank Raymond Stora for
drawing to my attention the work of S. Wu and for many helpful
discussions.

\vspace{1.5cm}

\section*{Appendix.}

It was already shown in \cite {STW94} that the Weil curvature takes
the form :

\begin{eqnarray}
\left( R_W^{eq}\right) _\mu ^\nu &=& \left( R ^{LC}
-i_\Sigma(\omega )R ^{LC} +\frac{i_\Sigma(\omega )
i_\Sigma(\omega )}{2}R ^{LC} + \frac{1}{2} D ^{LC}
\wedge \overline{\widetilde{\gamma}} \right. \\
&& \left.
-\frac{1}{2}i_\Sigma(\omega )
D ^{LC}\wedge \overline{\widetilde{\gamma}}
-\frac{1}{4} \widetilde{\psi}\widetilde{\psi} +\frac{1}{2}D
^{LC}\wedge
\overline{\widetilde{\Omega}}
 \right)_\mu ^\nu \nonumber
\end{eqnarray}

\noindent where:

\begin{eqnarray}
\overline{\widetilde{\gamma}}_\mu &=& \left( \delta g_{\rho \mu }
-l_\Sigma(\omega )g_{\rho \mu } \right)dx^\rho
=\widetilde{\gamma}_{\rho \mu }dx^\rho \\
\widetilde{\psi}_\mu ^\nu &=& g^{\rho \nu }\left( \delta g_{\rho \mu }
-l_\Sigma(\omega )g_{\rho \mu } \right)
=g^{\rho \nu }\left( \widetilde{\gamma}_{\rho \mu } \right)=
\left( g^{-1}\widetilde{\gamma} \right)_\mu ^\nu \\
\left( D ^{LC}\wedge \overline{\widetilde{\gamma}} \right)_\mu ^\nu
&=& g^{\rho \nu }\left(D _\rho ^{LC}
\overline{\widetilde{\gamma}}_\mu
-D _\mu ^{LC} \overline{\widetilde{\gamma}}_\rho \right)
\end{eqnarray}

Then, after a "straightforward" algebraic juggle, one finally
obtains :

\begin{eqnarray}
Q_4^0 &=& \dfrac{\varepsilon ^{\mu \nu \rho \sigma }}{\sqrt{{\bf
g}}} g_{\nu \lambda }g_{\sigma \chi } \left( R ^{LC}\right)_\mu
^\lambda
\wedge \left( R ^{LC}\right)_\rho ^\chi =E_\Sigma \\
\nonumber \\[2mm]
Q_3^1 &=& 2\dfrac{\varepsilon ^{\mu \nu \rho \sigma }}{\sqrt{{\bf g}}}
g_{\nu \lambda }g_{\sigma \chi }\left( R ^{LC}\right)_\mu
^\lambda\wedge \left(-i_\Sigma(\omega ) R^{LC}+\frac{1}{2} D ^{LC}
\wedge \overline{\widetilde{\gamma}}\right)_\rho ^\chi \\
\nonumber \\[2mm]
Q_2^2 &=& \dfrac{\varepsilon ^{\mu \nu \rho \sigma }}{\sqrt{{\bf
g}}} g_{\nu \lambda }g_{\sigma \chi } \left[ \left( i_\Sigma(\omega
) R^{LC}\right)_\mu ^\lambda \wedge \left( i_\Sigma(\omega )
R^{LC}\right)_\rho ^\chi -2\left( i_\Sigma(\omega )
R^{LC}\right)_\mu
^\lambda \wedge \left(D ^{LC} \wedge \overline{\widetilde{\gamma}}
\right)_\rho ^\chi \right. \\
&& \phantom{\dfrac{\varepsilon ^{\mu \nu \rho \sigma }}{\sqrt{{\bf
g}}} g_{\nu \lambda }} \left. + \left( D ^{LC}
\wedge\overline{\widetilde{\gamma}}\right)_\mu
^\lambda \wedge \left( D ^{LC} \wedge\overline{\widetilde{\gamma}}
\right)_\rho^\chi \right. \nonumber \\
&& \phantom{\dfrac{\varepsilon ^{\mu \nu \rho \sigma }}{\sqrt{{\bf
g}}} g_{\nu \lambda }} + \left. \left( R ^{LC}\right)_\mu ^\lambda
\wedge
\left( i_\Sigma(\omega )i_\Sigma(\omega )R ^{LC} - i_\Sigma(\omega )
\left( D ^{LC} \wedge \overline{\widetilde{\gamma}}\right) - \frac{1}{2}
\widetilde{\psi}\widetilde{\psi} - D ^{LC}\wedge
\overline{\widetilde{\Omega}}\right)_\rho ^\chi \right] \nonumber \\
\nonumber \\[2mm]
Q_1^3 &=& \dfrac{\varepsilon ^{\mu \nu \rho \sigma }}{\sqrt{{\bf g}}}
g_{\nu \lambda }g_{\sigma \chi } \left( i_\Sigma(\omega )
i_\Sigma(\omega )R ^{LC} -i_\Sigma(\omega ) D ^{LC} \wedge
\overline{\widetilde{\gamma}} -\frac{1}{2}
\widetilde{\psi}\widetilde{\psi} -D ^{LC}\wedge
\overline{\widetilde{\Omega}}\right)_\mu ^\lambda \\
&& \phantom{\dfrac{\varepsilon ^{\mu \nu \rho \sigma }}{\sqrt{{\bf
g}}} g_{\nu \lambda }} \ \wedge \left( -i_\Sigma(\omega )
R^{LC}+\frac{1}{2} D^{LC} \wedge \overline{\widetilde{\gamma}}
\right)_\rho ^\chi \nonumber \\
\nonumber \\[2mm]
Q_0^4 &=& \dfrac{\varepsilon ^{\mu \nu \rho \sigma }}{4\sqrt{{\bf g}}}
g_{\nu \lambda }g_{\sigma \chi } \left( i_\Sigma(\omega )
i_\Sigma(\omega )R ^{LC} -i_\Sigma(\omega ) D^{LC} \wedge
\overline{\widetilde{\gamma}} -\frac{1}{2}
\widetilde{\psi}\widetilde{\psi} -D ^{LC}\wedge
\overline{\widetilde{\Omega}}\right)_\mu ^\lambda \\
&& \phantom{\dfrac{\varepsilon ^{\mu \nu \rho \sigma }}{4\sqrt{{\bf
g}}} g_{\nu \lambda }} \ \wedge \left( i_\Sigma(\omega )
i_\Sigma(\omega )R^{LC}-i_\Sigma(\omega ) D ^{LC} \wedge
\overline{\widetilde{\gamma}} -\frac{1}{2}
\widetilde{\psi}\widetilde{\psi} -D ^{LC}\wedge
\overline{\widetilde{\Omega}}\right)_\rho ^\chi \nonumber
\end{eqnarray}

Finally, the G's are obtained by replacing $\dfrac{\varepsilon ^{\mu
\nu\rho\sigma }}{\sqrt{{\bf g}}} g_{\nu \lambda }g_{\sigma \chi }$ in
the Q's by $\left(\delta^{\mu}_{\lambda} \delta^{\rho}_{\chi} -
\delta^{\mu}_{\chi} \delta^{\rho}_{\lambda} \right)$.

%\newpage

\end{document}